# Evidence of shallow bandgap in ultra-thin 1T'-MoTe$_2$ via infrared spectroscopy

Jin Cheol Park[1,2,§], Eilho Jung[3,§], Sangyun Lee[2], Jungseek Hwang[3,*], Young Hee Lee[1,2,3,*]

[1]Department of Energy Science, Sungkyunkwan University, Suwon 16419, Republic of Korea.

[2]Center for Integrated Nanostructure Physics (CINAP), Institute for Basic Science (IBS), Suwon 16419, Republic of Korea.

[3]Department of Physics, Sungkyunkwan University, Suwon 16419, Republic of Korea.

[*]Corresponding author, email: jungseek@skku.edu (J. Hwang), leeyoung@skku.edu (Y.H. Lee)

[§]These authors (J.C.P. and E.J.) contributed equally to this work.

**ASTRACT**

Although van der Waals (vdW) layered $MoS_2$ shows the phase transformation from the semiconducting 2H-phase to the metallic 1T-phase through chemical lithium intercalation, vdW $MoTe_2$ is thermodynamically reversible between the 2H- and 1T'-phases, and can be further transformed by energetics, laser irradiation, strain or pressure, and electrical doping. Here, thickness- and temperature-dependent optical properties of 1T'-$MoTe_2$ thin films grown by chemical vapor depsition are investigated *via* Fourier-transformed infrared spectroscopy. An optical gap of 28 $\pm$2 meV in a 3-layer (or 2 nm) thick 1T'-$MoTe_2$ is clearly observed at a low temperature region below 50K. No discernible optical bandgap is observed in samples thicker than ~4 nm. The observed thickness-dependent bandgap results agree with the measured dc resistivity data; the thickness-dependent 1T'-$MoTe_2$ clearly demonstrates the metal-semiconductor transition at a crossover below the 2 nm-thick sample.

Two-dimensional (2D) van der Waals (vdW) layered materials including graphene, hexagonal boron nitride (h-BN), and transition metal dichalcogenides (TMDs) have demonstrated numerous intriguing physical phenomena including spin-orbit coupling, strong Coulomb interaction, large exciton binding energy, bandgap renormalization, direct-to-indirect bandgap transition, topological insulator, valleytronics, and metal-insulator transition [1-6]. Among 2D vdW layered materials, TMD materials have especially shown various phase-dependent crystal forms including 2H-, 1T-, 1T'-, rhombohedral 3R- and orthorhombic $T_d$-phases [7-9]. Various semiconducting 2H-phases have received considerable attention because of their fascinating electrical and optical properties, e.g., indirect-to-direct bandgap crossover, large exciton binding energy, and spin-orbit splitting of excitons [10,11], together with the recently observed semi-metallic states (1T-, 1T'-, and $T_d$-phases) for charge density wave, unsaturated and giant magnetoresistance and Weyl semimetal (WSM) [12,13].

1T'-Molybdenum ditelluride ($MoTe_2$) has been suggested as a type II WSM candidate [14-17], and recently proposed as a Quantum spin Hall (QSH) insulator [18]. The presence of bandgap of ~0.1 eV, its value and sensitivity to temperature are primary concern in the field of QSH-based devices for practical applications [19]. For practical applications, information of optical (or electronic) properties of thin film 1T'-$MoTe_2$ is important. Therefore, one needs to study thickness- and temperature-dependent optical properties of thin film 1T'-$MoTe_2$ to search for a bandgap opening.

Here, we report the thickness- and temperature-dependent optical properties of CVD-grown 1T'-$MoTe_2$ thin films *via* Fourier-transform infrared spectroscopy (FTIR). Temperature-dependent optical conductivity spectra of 1T'-$MoTe_2$ films are observed for three different thicknesses of 2, 4, and 10 nm. We directly observe a narrow bandgap of 28 ±2 meV in a 3 layer (or 2 nm) thin 1T'-MoTe$_2$ at temperature below 50K; this narrow bandgap is obscured by thermally excited charge carriers at temperatures above 80K. Furthermore, these observations are congruent with the electrically measured dc resistivity data of the thin film samples. Moreover, we discover the phase transition from the 1T'- to T$_d$-phase below 250K for a thickness of above 10 nm.

A schematic of a two-zone CVD system was adopted to separately control the temperature and heating rate of both the Te supplying ($T_1$) and synthesis zones ($T_2$) (Fig. 1a and Fig. S1, Supplemental Material; details are provided in Methods). To characterize the CVD-grown 1T'-MoTe$_2$ films, Raman spectroscopy are performed with a 532 nm wavelength laser. Two important vibration modes, A$_g$ (258–265 cm$^{-1}$) and B$_g$ (162–163 cm$^{-1}$) are observed in three different thickness samples. The position of B$_g$ mode does not vary with thicknesses, while the position of A$_g$ mode exhibits a blue-shift from 258.8 to 265.0 cm$^{-1}$ with decreasing thickness (Fig. 1b), which agrees with previous reports [20]. This implies that the samples are fully relaxed after transfer of the samples. To confirm the presence of strain, we measured Raman spectra (B$_g$ mode) of 1T'-MoTe$_2$ film samples before and after the transfer. No significant difference between two sets of B$_g$ modes is observed (Fig. S2). Additionally, X-ray diffraction (XRD), transmission electron microscopy (TEM) and scanning transmission electron microscopy (STEM) were used to characterize the thin film samples in previous report [21].

1T'-MoTe$_2$ films were prepared on a centimeter-scale double-side polished undoped Si substrate for optical study (Fig. 1c). The transmittance spectra were measured with respect to the substrate ($T(\omega) = \frac{T_{film/substrate}}{T_{substrate}}$), where $T_{film/substrate}$ and $T_{substrate}$ are the transmittance spectra of the sample/substrate and substrate, respectively. We show the measured transmittance spectra ($T(\omega)$) of the three 1T'-MoTe$_2$ samples with 2, 4, and 10 nm thicknesses at various temperatures (Fig. 1c); transmittance increases in thinner 1T'-MoTe$_2$ samples. The transmittance is reduced in the far-IR region below 400 cm$^{-1}$ for thicknesses of 4 and 10 nm, showing a metallic behavior with a Drude component. In the same region, the transmittance increases at a thickness of 2 nm particularly in the low temperature region below 200K.

To further understand the distinct behaviors between a 2 nm-thin sample and thicker samples in the far-IR region, the optical conductivity is extracted from the observed transmittance spectra via the Tinkham formula, which can be written as, $T(\omega) = \frac{1}{\left|1+\tilde{\sigma}(\omega)\, d\frac{Z_0}{n_s+1}\right|^2}$, where $T(\omega)$ is the transmittance with respect to the substrate, $\tilde{\sigma}(\omega) = \sigma_1(\omega) + i\sigma_2(\omega)$ is the complex optical conductivity, $n_s$ is the index of refraction of the substrate, and $Z_0$ (= 377 Ω) is the impedance of the vacuum [22-25]. For very thin films on a thick substrate ($d \ll d_{sub}, d \ll \lambda$, and $d \ll \delta$, where $d$, $d_{sub}$, $\delta$, and $\lambda$ are the thicknesses of the film and the substrate, the skin depth (for metallic films), and the wavelength, respectively) the real part of the optical conductivity can be rewritten as, $\sigma_1(\omega) \cong \frac{n_s+1}{Z_0 d}\left(\sqrt{\frac{1}{T(\omega)}} - 1\right)$. Here we note that the contribution of the imaginary part of the optical conductivity is negligible for metals in low frequency region [25]. To further investigate the electronic structure, we used the Drude-Lorentz model; intraband and interband optical transitions can be described by their respective Drude and Lorentz components. The real part of the optical conductivity can be described in the Drude-Lorentz model as, $\sigma_1(\omega) = \frac{1}{4\pi}\left[\frac{\omega_{pD}^2}{\tau(\omega^2+\tau^{-2})} + \sum_j \frac{\gamma_j \omega^2 \Omega_j^2}{\left(\omega_j^2-\omega^2\right)^2+\gamma_j^2\omega^2}\right]$, where $\omega_{pD}$ and $\tau^{-1}$ are the plasma frequency and impurity scattering rate of the Drude mode, respectively; $\Omega_j$, $\omega_j$, and $\gamma_j$ are the plasma frequency, resonance frequency, and damping parameter of the *j*th Lorentz mode, respectively. Moreover, one can obtain the dc conductivity from the zero-frequency limit of the optical conductivity as $\sigma_1(0) = \frac{\omega_{pD}^2}{4\pi\tau^{-1}}$, which can be described by the Drude plasma frequency and scattering rate. The optical conductivity contains information of the electronic (band) structure; intraband transitions in the partially filled bands are contributed from unbound electrons (or the Drude component), and interband transitions from the occupied bands to the unoccupied bands are contributed from bounded electrons (or the Lorentz components). The real part of the optical conductivity in the infrared region can be fitted using the Drude-Lorentz model.

The optical conductivities are extracted from the measured transmittance spectra of three different samples with thicknesses of 10, 4, and 2 nm at 8K and 300K (inset) (Fig. 2, with more temperature points in Fig. S3, S4, S5 and S6, Supplemental Material). Drude and Lorentz components are used for analyzing the optical conductivities in an entire range of frequencies up to 1500 cm$^{-1}$. The fitted results are shown with the dashed red lines, which are composed of one Drude component (the solid wine line) and a pair of Lorentz components ($L_1$ and $L_2$, the solid blue line denotes the optical conductivity without the Drude component) in Fig. 2. Drude components below 400 cm$^{-1}$ are clearly observed in the optical conductivity at thicknesses of 10 and 4 nm (Fig. 2a and 2b). In contrast, the Drude component is absent in the 2 nm-thick sample at 8K (Fig. 2c). From the optical conductivity in the 2 nm-thick sample, we extract an optical gap ($E_g$) of 28 $\pm$2 meV at 8K. The error in the gap is roughly estimated from uncertainty of the linear extrapolation line (blue dashed line). The optical gap is closed at temperatures above 80K.

Here we need to discuss about applicability of the Tinkham formula used for metallic samples, $\sigma_1(\omega) \cong \frac{n_s+1}{Z_0 d}\left(\sqrt{\frac{1}{T(\omega)}} - 1\right)$, to a spectrum with an energy gap because in this case the imaginary part of the conductivity ($\sigma_2(\omega)$) is not guaranteed to be negligibly small. However, it has been shown that for very thin semiconducting samples the imaginary part can be ignored [26]. Further detailed discussion on this issue is given in the Supplemental Material (Fig. S7).

The fitting parameters of the Drude component are the plasma frequency, $\omega_{pD}$ and the impurity scattering rate, $\tau^{-1}$ (Fig. 3a, 3b and 3c). For the 10 nm-thick 1T'-MoTe$_2$ sample, the Drude plasma frequency ($\omega_{pD}$) is slightly enhanced from 4400 to 4700 cm$^{-1}$ below 250K as the temperature decreases. Such an gradual rise of $\omega_{pD}$ at 250K is associated with an emergence of a Weyl semimetallic state in the band structure, which takes place along with the structural phase transition from 1T'- to T$_d$-phase at approximately 200K [8,27,28]. Similar behaviors are observed with charge carrier densities varying from $1.49 \times 10^{13}$ to $1.70 \times 10^{13}$ cm$^{-2}$ (Fig. S8, Supplemental Material). In contrast, the $\tau^{-1}$ decreases linearly from 230 cm$^{-1}$ at 150K to 192.5 cm$^{-1}$ at 8K, i.e. the Drude component becomes sharper, which is a characteristic of typical metals. For the 4 nm-thick 1T'-MoTe$_2$, the plasma frequency (or charge carrier density) and scattering rate monotonically decrease with decreasing temperature and become more prominent with monotonic frequency behavior in 2 nm-thick MoTe$_2$; this indicates that the T$_d$-phase appears in bulk but vanishes in thin MoTe$_2$ film.

We describe the thermally induced dc conductivity in terms of the optical bandgap ($E_g$) using the Arrhenius plot, $\sigma_{dc} = \sigma_0 \cdot \exp\left(-\frac{E_g}{2 \cdot k_B T}\right)$, where $\sigma_{dc}$ is the dc conductivity, $\sigma_0$ is the pre-exponential factor, $k_B$ is the Boltzmann constant, and $T$ is the absolute temperature [29,30]. The pre-factor (1/2) in the exponent appears due to the possible distributions of electrons in the conduction band, independent of those of the valence band holes. The dc conductivity for 2 nm-thick 1T'-MoTe$_2$ sample yields $\sigma_0$ = 1300 $\Omega^{-1}$cm$^{-1}$ and $E_g$ = 36 $\pm$4 meV (Fig. 3d). The error in the gap is estimated by adjusting the fitting parameters for giving similar fitting quality. This is consistent with the optical gap obtained from the optical conductivity (see Fig. 2c). Here we note that the optical gap (28 $\pm$2 meV) directly obtained from optical conductivity is more reliable compared with the one (36 $\pm$4 meV) obtained from the thermally induced dc conductivity.

We now demonstrate the experimental optical gap in our temperature-dependent spectroscopy. Our results are consistent with a finite bandgap predicted by the theoretical band structure of monolayer 1T'-MoTe$_2$ (Fig. 3e). As the temperature increases, electrons in the valence band can be thermally excited to the conduction band, thus contributing to the conductivity. At temperatures above 80K at 2 nm-thick 1T'-MoTe$_2$, the optical conductivity includes a Drude component from these thermally excited electrons. One may concern static doping, which may be caused by possible impurities, in the samples prepared by CVD. We do not directly check whether our thin film samples with thicknesses of 4 nm and 10 nm are statically doped or not. However, we can tell that the 2 nm-thick sample is not statically doped because we can clearly observe a small energy gap (28 $\pm$2 meV). Furthermore, since all samples were prepared by the same CVD method, we expect that all samples prepared by our CVD method are free from static doping.

Finally, we display the optically extracted dc resistivity data of our three 1T'-MoTe$_2$ thin film samples with their corresponding temperatures (Fig. 4a). The dc resistivity data of 10 and 4 nm-thick 1T'-MoTe$_2$ samples show metallic behaviors. Meanwhile, a semiconducting behavior emerges at 2 nm-thick 1T'-MoTe$_2$ sample. This is a clear demonstration of a thickness-induced metal-semiconductor transition between 2 and 4 nm. We also measured the dc resistivity data using the van der Pauw method (Details are described in Methods) [31]. Similar metal-semiconductor turnover is observed, although the absolute resistivity value varies from the optical gap approach (Fig. 4b). The optically estimated resistivity may contain large uncertainty due to unavailable frequency below ~70 cm$^{-1}$ compared to directly measured dc resistivity, the conductivity fitted from Drude model may cause to overetimate at zero frequency limit.

We note that the small bandgap of 28 ±2 meV in 2 nm-thick 1T'-MoTe$_2$ is different from previous reported 60 meV in exfoliated single-crystal 1T'-MoTe$_2$ [20]. The previously measured bandgap could be a mixed (or averaged) bandgap of 1~10-layered 1T'-MoTe$_2$. It is of note that our samples grown by CVD might contain some amount of defects, which may smear the size of the bandgap. To estimate the crystal quality, we provide the TEM images of top and side views of multilayer 1T'-MoTe$_2$ film, indicating the presence of random Te vacancies (Fig. S9).

In conclusion, we have successfully studied the bandgap evolution of 1T'-MoTe$_2$ with respect to both thickness and temperature. The Drude component becomes dominant at bulk samples (4 and 10 nm samples) in all temperature regions and at a thickness of 2 nm above 80K. The T$_d$-phase has appeared in the 10 nm-thick sample below 250K. The 2 nm-thick 1T'-MoTe$_2$ sample reveals an optical bandgap of 28 ±2 meV at 8K that vanishes at temperatures above 80K. The vanished bandgap at high temperatures can be attributed to the thermally excited electrons from the valence to the conduction bands. Moreover, we have independently extracted a bandgap of 36 ±4 meV from the thermally induced dc conductivity. No bandgap opening is observed at 4 and 10 nm-thick samples at all selected temperatures between 8 and 350K. Thickness-dependent MoTe$_2$ clearly demonstrates metal-semiconductor transition at a crossover below the 2 nm-thick sample. In addition, both the dc resistivity extracted from the optical conductivity spectra and that which was directly measured by the van der Pauw method show similar temperature-dependent behaviors [31]. Thus, our approach can be applied to other semi-metallic two-dimensional layered materials for investigating their unique optical and electronic properties.

**Methods**

A. Synthesis of 1T'-MoTe$_2$ films

To synthesize semi-metallic 1T'-MoTe$_2$ thin film, Mo thin films with thicknesses of 0.5 nm, 1.3 nm, and 4 nm were deposited on a 300 nm-thick SiO$_2$/Si substrate via a DC sputter. In a two-zone chemical vapor deposition (CVD) system, a Te pellet of 0.5 g (Sigma-Aldrich) was placed in the first furnace zone and the Mo-deposited substrate was placed in the second furnace zone. To control the tellurization rate, the temperatures of the Te zone ($T_1$) and Mo film zone ($T_2$) were controlled separately. Before growing the 1T'-MoTe$_2$ film, the whole CVD system was purged with 1000 sccm of argon (Ar) gas for 30 min to create an oxygen-free and Ar environment. During the growth process, Ar and hydrogen (H$_2$) gases were flown with rates of 100 and 20 sccm, respectively. $T_1$ and $T_2$ started heating up simultaneously with a ramping rate of $T_1$ reaching 555 °C first in 13 min and then $T_2$ reaching 650 °C in 15 min. When the $T_2$ temperature reached 650 °C, growth was carried out for 30 min under atmospheric pressure. After this growth, $T_1$ was cooled rapidly by opening the chamber; the $T_2$ chamber was opened 1 min after from opening of the $T_1$ chamber to provide further cooling. During the cooling process, 500 sccm Ar and 20 sccm H$_2$ gases were introduced to remove the reactants. Finally, 2, 4, and 10 nm-thick 1T'-MoTe$_2$ films were produced.

B. Transfer of 1T'-MoTe$_2$ onto undoped Si substrate

To measure the FTIR spectra, the samples were transferred to undoped Si substrate using the conventional poly methyl methacrylate (PMMA) method [21,32]. To detach the grown 1T'-MoTe$_2$ film from SiO$_2$/Si substrate on CVD, the PMMA-coated 1T'-MoTe$_2$ film was floated on a buffered oxide etch (B.O.E., 1178-03, J.T. Baker) for 20 min. The PMMA/1T'-MoTe$_2$ film was then rinsed with distilled water several times after the SiO$_2$ part was completely etched by the B.O.E. solution. The resulting film was transferred onto an undoped Si substrate. After drying the film, the PMMA layer was eliminated by applying acetone and isopropyl alcohol (IPA).

C. Fourier-transform infrared spectroscopy (FTIR)

FTIR is an experimental technique to obtain the power spectrum in a wide infrared range from far- to near-infrared. To measure the transmittance spectra of our samples we used a commercial FTIR spectrometer—Bruker Vertex 80v. The Vertex 80v is a vacuum-type spectrometer that is free from moisture and air absorptions. The spectrometer consists of three important components: a source, beam splitter, and detector. The configuration of the three components consists of a mercury arc lamp source, 6-micron multilayer Mylar beam splitter, and 4K-bolometer detector for the far-infrared (50–700 cm$^{-1}$ or 0.006–0.086 eV) range, a Globar lamp source, KBr beam splitter, and MCT (or RT-DTGS) detector for mid-infrared (400–7000 cm$^{-1}$ or 0.05–0.868 eV) range, and a tungsten lamp source, CaF$_2$ beam splitter, and InGaAs detector for near-infrared (4000–12000 cm$^{-1}$ or 0.5–1.5 eV). We used a commercial cold finger-type ARS optical cryostat and a commercial temperature controller (Lakeshore 325) to control the sample temperature both below and above room temperature. We used liquid helium as a coolant and were able to control the sample temperature from 8 to 350K.

D. The dc-transport resistivity measurement

The electrical resistivity was measured by the standard van der Pauw method. Platinum wires were attached by silver paste on each corner of the samples to provide a contact to the samples. The samples were cooled down by closed cycle cryostat and measured using a Lake Shore Cryotronics® model 370 AC resistance bridge.

**ACKNOWLEDGEMENTS**

This work was supported by the Institute for Basic Science of Korea (IBS-R011-D1). J.H. acknowledges financial support from the National Research Foundation of Korea (NRF-2017R1A2B4007387).

**FIGURES AND FIGURE CAPTIONS**

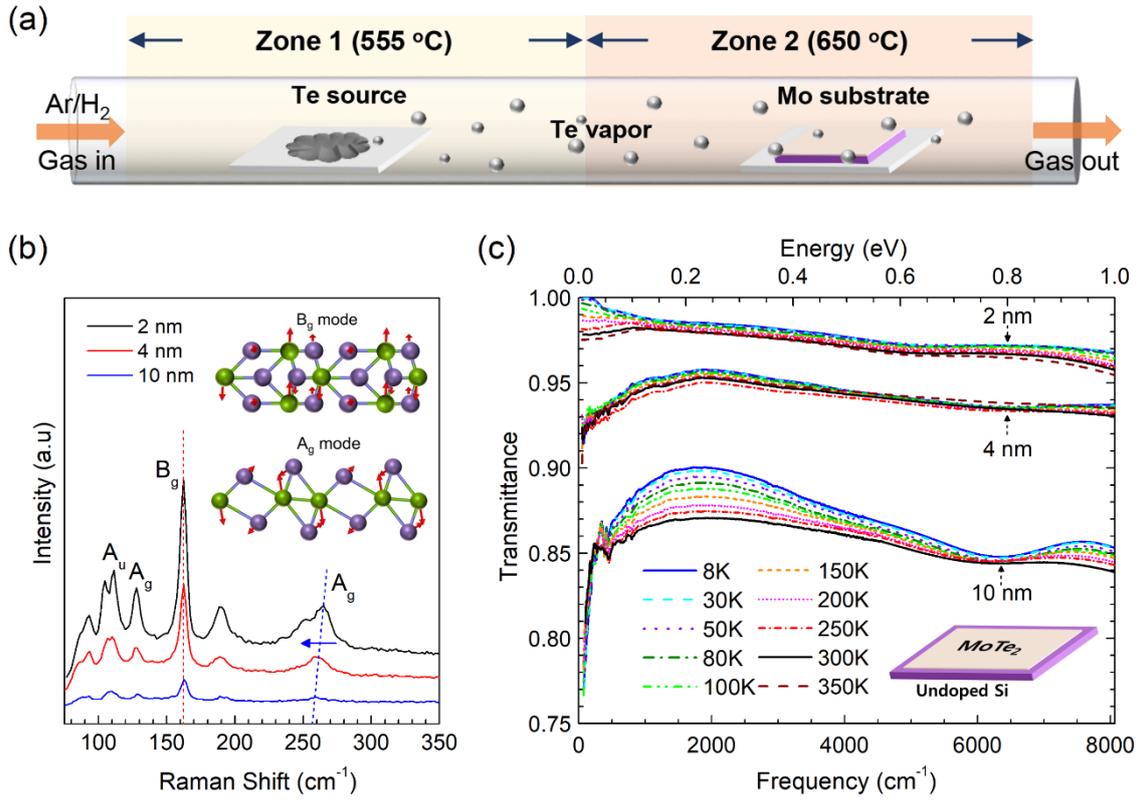

FIG. 1. Synthesis of 1T'-MoTe$_2$ thin films and transmittance spectra. (a) A schematic of a 2-zone chemical vapor deposition system to synthesize the 1T'-MoTe$_2$ film on SiO$_2$/Si substrate. (b) Thickness-dependent Raman spectra for 1T'-MoTe$_2$ with three different thicknesses (2 nm, 4 nm, and 10 nm). (c) Measured transmittance spectra for 1T'-MoTe$_2$ films at various temperatures between 8K and 350K.

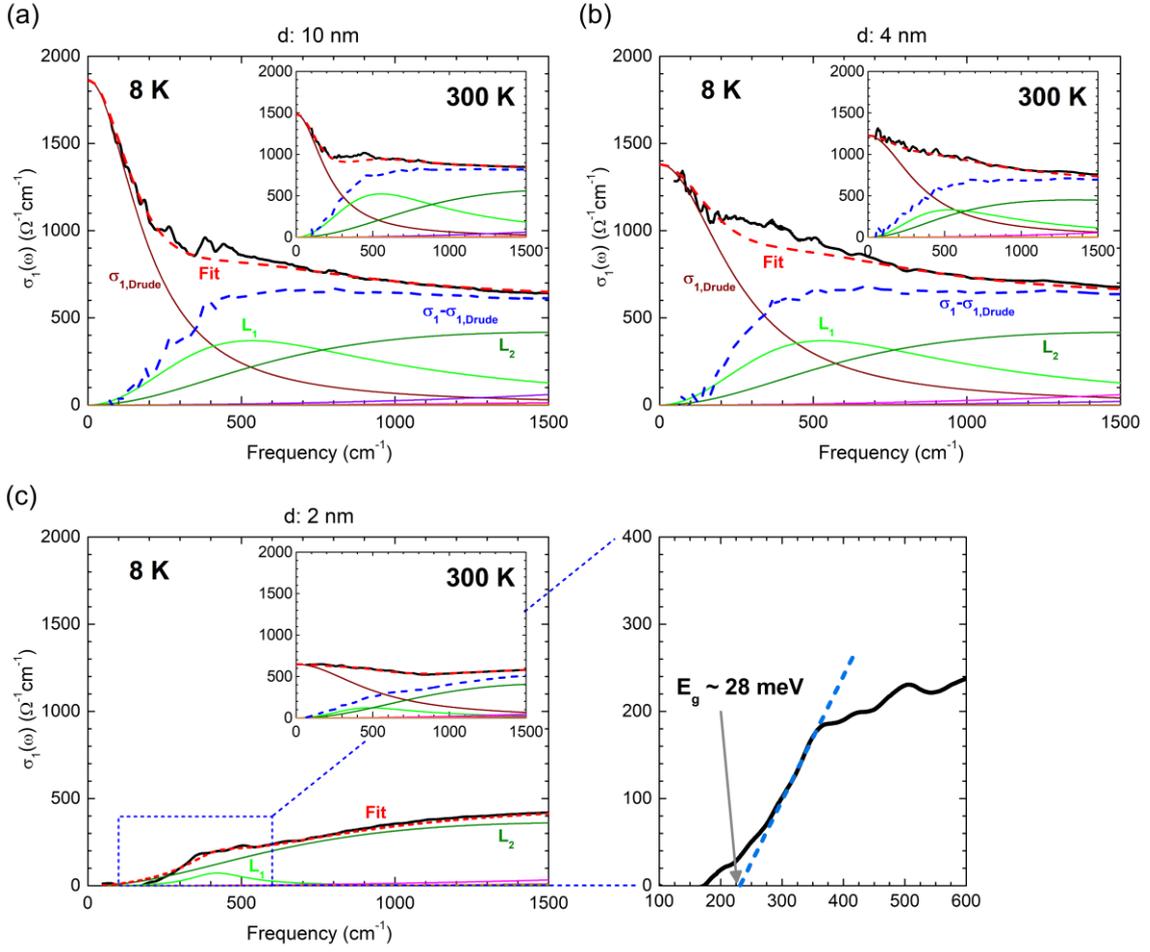

FIG. 2. Thickness- and temperature-dependent optical conductivity of 1T'-MoTe$_2$ films. The real part of optical conductivity for (a) 10 nm-thick 1T'-MoTe$_2$, (b) 4 nm-thick 1T'-MoTe$_2$, and (c) 2 nm-thick 1T'-MoTe$_2$ at 8K. Each inset shows the corresponding real part of optical conductivity at 300K.

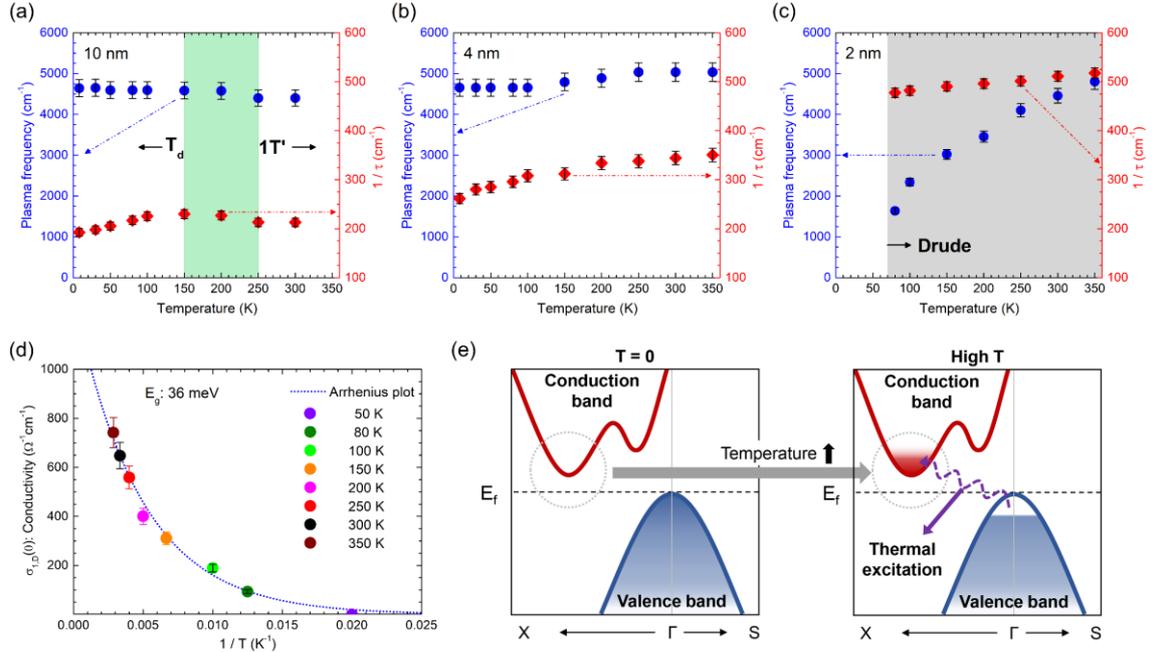

FIG. 3. Drude fitting parameters of 1T'-MoTe$_2$ films. The plasma frequency (blue triangle) and scattering rate (red triangle) of the Drude component obtained from the Drude-Lorentz model fit as a function of temperature in (a) 10 nm-thick, (b) 4 nm-thick, and (c) 2 nm-thick 1T'-MoTe$_2$. (d) The dc conductivity data obtained from a zero-frequency extrapolation of the optical conductivity and the fit (blue dotted line) as a function of $1/T$ at 2 nm-thick 1T'-MoTe$_2$. (e) Schematics of electronic band structures near the Fermi level at T = 0 (left) and thermally promoted excitations at a finite temperature (right).

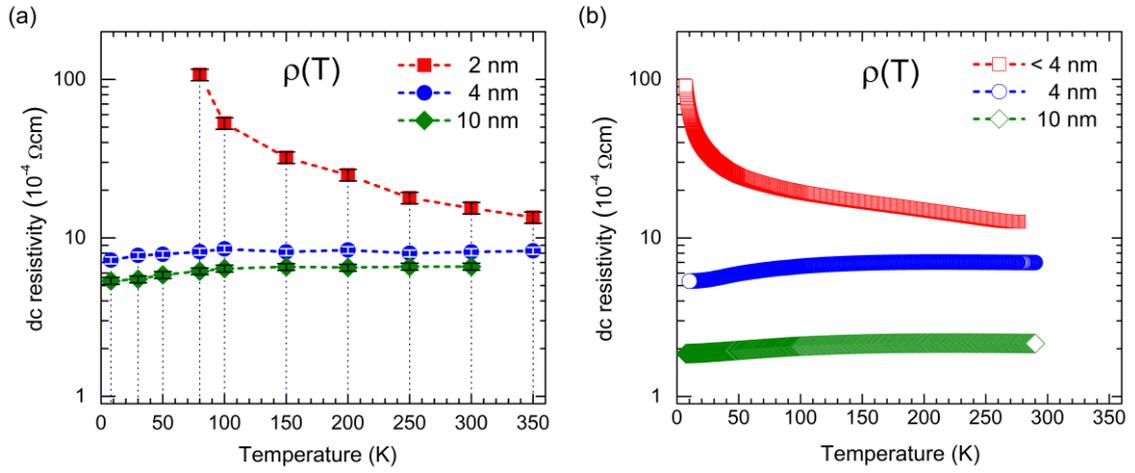

FIG. 4. Comparison of optical dc resistivity with directly measured dc-transport resistivity. (a) Optically obtained dc resistivity data of 1T'-MoTe$_2$ films with three different thicknesses as a function of temperature. (b) The directly measured dc resistivity data of the 1T'-MoTe$_2$ films as a function of temperature.